\documentclass[12pt]{llncs}
\usepackage{fullpage}

\PassOptionsToPackage{hyphens}{url}\usepackage{hyperref}
\usepackage{framed}
\usepackage{mdframed}
\usepackage{caption}

\usepackage{balance}
\usepackage{mathtools}
\usepackage{seqsplit}
\usepackage{amsmath}
\usepackage{algorithmic}
\usepackage{array}
\usepackage{wrapfig}
\usepackage{graphicx}
\usepackage{times}
\usepackage{setspace}
\usepackage[caption=false]{subfig}
\usepackage{graphicx}
\usepackage{enumitem}
\usepackage[boxed,ruled,linesnumbered]{algorithm2e}
\usepackage[none]{hyphenat}
\usepackage{theorem}
\usepackage{csquotes}

\newcommand{\B}{\vspace*{-\smallskipamount}}
\newcommand{\BB}{\vspace*{-\medskipamount}}
\newcommand{\BBB}{\vspace*{-\bigskipamount}}

\begin{document}
\title{Exploiting Data Sensitivity on Partitioned Data}
\author{Sharad Mehrotra, Kerim Yasin Oktay, \and Shantanu Sharma}
\institute{Department of Computer Science, University of California, Irvine, USA.\\ \email{sharad@ics.uci.edu, shantanu.sharma@uci.edu}.
\thanks{\textbf{
This chapter will appear in the book titled ``From Database to Cyber Security: Essays Dedicated to Sushil Jajodia on the Occasion of His 70th Birthday.'' For the final version, please check \url{https://link.springer.com/book/10.1007\%2F978-3-030-04834-1}
The full approaches proposed in this chapter may be found in~\cite{DBLP:conf/icde/OktayKM17,TR-2018}}. This material is based on research sponsored by DARPA under agreement number FA8750-16-2-0021. The U.S. Government is authorized to reproduce and distribute reprints for Governmental purposes notwithstanding any copyright notation thereon. The views and conclusions contained herein are those of the authors and should not be interpreted as necessarily representing the official policies or endorsements, either expressed or implied, of DARPA or the U.S. Government. This work is partially supported by NSF grants 1527536 and 1545071.}}

\maketitle
\thispagestyle{empty}

\begin{abstract}
Several researchers have proposed solutions for secure data outsourcing on the public clouds based on encryption, secret-sharing, and trusted hardware. Existing approaches, however, exhibit many limitations including high computational complexity, imperfect security, and information leakage. This chapter describes an emerging trend in secure data processing that recognizes that an entire dataset may not be sensitive, and hence, non-sensitivity of data can be exploited to overcome some of the limitations of existing encryption-based approaches. In particular, data and computation can be partitioned into sensitive or non-sensitive datasets -- sensitive
data can either be encrypted prior to outsourcing or stored/processed locally on trusted servers. The non-sensitive dataset, on the other hand, can be outsourced and processed in the cleartext.
 While partitioned computing can bring new efficiencies since it does not incur (expensive) encrypted data processing costs on non-sensitive data, it can lead to  information leakage. We study partitioned computing in two contexts - first,  in the context of the hybrid cloud where  local resources are integrated with public cloud resources to form an effective and secure storage and computational platform for enterprise data. In the hybrid cloud, sensitive data is stored on the private cloud to prevent leakage and a computation is partitioned between private and public clouds. Care must be taken that the public cloud cannot infer any information about sensitive data from inter-cloud data access during query processing. We then consider partitioned computing in a public cloud only setting, where sensitive data is encrypted before outsourcing. We  formally define a {\em partitioned security} criterion that any approach to partitioned computing on public clouds must ensure in order to not introduce any new vulnerabilities to the existing secure solution. We sketch out an approach to secure partitioned computing that we refer to as {\em query binning} (QB) and show how QB can be used to support selection queries. We evaluate conditions under which partitioned computing approaches such as QB can improve the performance of cryptographic approaches that are prone to size, frequency-count, and workload attacks.
\end{abstract}

\section{Introduction}
\label{sec:introduction}
Organizations today collect and store a large volume of data, which is analyzed for diverse purposes. However, in-house computational capabilities of organizations may become obstacles for storing and processing data. Many \emph{untrusted cloud computing} platforms (\textit{e}.\textit{g}., Amazon AWS, Google App Engine, and Microsoft Azure) offer database-as-a-service using which data owners, instead of purchasing, installing, and running data management systems locally, can outsource their databases and query processing to the cloud. Such cloud-based services available using the pay-as-you-go model offers significant advantages to both small, medium and at times large organizations. The numerous benefits of public clouds impose significant security and privacy concerns related to sensitive data storage (\textit{e}.\textit{g}., sensitive client information, credit card, social security numbers, and medical records) or the query execution. The untrusted public cloud may be an \emph{honest-but-curious} (or passive) adversary, which executes an assigned job but tries to find some meaningful information too, or a malicious (or active) adversary, that may tamper the data or query.
Such concerns are not a new revelation -- indeed, they were identified as a key impediment for organizations adopting the database-as-as-service model in early work on data outsourcing \cite{DBLP:conf/sigmod/HacigumusILM02,DBLP:conf/icde/HacigumusMI02}. Since then, security/confidentiality challenge has been extensively studied in both the cryptography and database literature, which has resulted in many techniques to achieve \emph{data privacy}, \emph{query privacy}, and \emph{inference prevention}. Existing work can loosely be classified into the following three categories:
\begin{enumerate}
  \item \textbf{Encryption based techniques.} \textit{E}.\textit{g}., order-preserving encryption~\cite{DBLP:conf/sigmod/AgrawalKSX04}, deterministic encryption (Chapter 5 of~\cite{DBLP:books/cu/Goldreich2004}), homomorphic encryption~\cite{gentry2009fully}, bucketization~\cite{DBLP:conf/sigmod/HacigumusILM02}, searchable encryption~\cite{DBLP:conf/sp/SongWP00}, private informational retrieval (PIR)~\cite{DBLP:journals/jacm/ChorKGS98}, practical-PIR (P-PIR)~\cite{DBLP:journals/iacr/WangDDB06}, oblivious-RAM (ORAM)~\cite{DBLP:conf/stoc/Goldreich87}, oblivious transfers (OT)~\cite{DBLP:journals/iacr/Rabin05}, oblivious polynomial evaluation (OPE)~\cite{DBLP:journals/siamcomp/NaorP06}, oblivious query processing~\cite{DBLP:conf/icdt/ArasuK14}, searchable symmetric encryption~\cite{DBLP:journals/jcs/CurtmolaGKO11}, and distributed searchable symmetric encryption (DSSE)~\cite{DBLP:conf/ctrsa/IshaiKLO16}.
  \item \textbf{Secret-sharing~\cite{DBLP:journals/cacm/Shamir79} based techniques.} \textit{E}.\textit{g}., distributed point function~\cite{DBLP:conf/eurocrypt/GilboaI14}, function secret-sharing~\cite{DBLP:conf/eurocrypt/BoyleGI15}, functional secret-sharing~\cite{DBLP:conf/tcc/KomargodskiZ16}, accumulating-automata~\cite{DBLP:conf/ccs/DolevGL15,DBLP:conf/dbsec/DolevL016}, \textsc{Obscure}~\cite{citing_obscure}, and others~\cite{DBLP:journals/isci/EmekciMAA14,DBLP:conf/fc/LueksG15,DBLP:journals/iacr/LiMD14}.
  \item \textbf{Trusted hardware-based techniques.} They are either based on a secure coprocessor or Intel SGX, \textit{e}.\textit{g}.,~\cite{DBLP:conf/cidr/ArasuBEKKRV13,DBLP:journals/pvldb/BajajS13}. The secure coprocessor and Intel SGX~\cite{DBLP:journals/iacr/CostanD16} allow decrypting data in a secure area and perform some computations. 
\end{enumerate}
\mdfdefinestyle{mystyle}{leftmargin=0.5cm,rightmargin=0.5cm,linecolor=black}
\begin{figure}[!h]
\begin{mdframed}[style=mystyle]
\begin{center}
\includegraphics[scale=0.5]{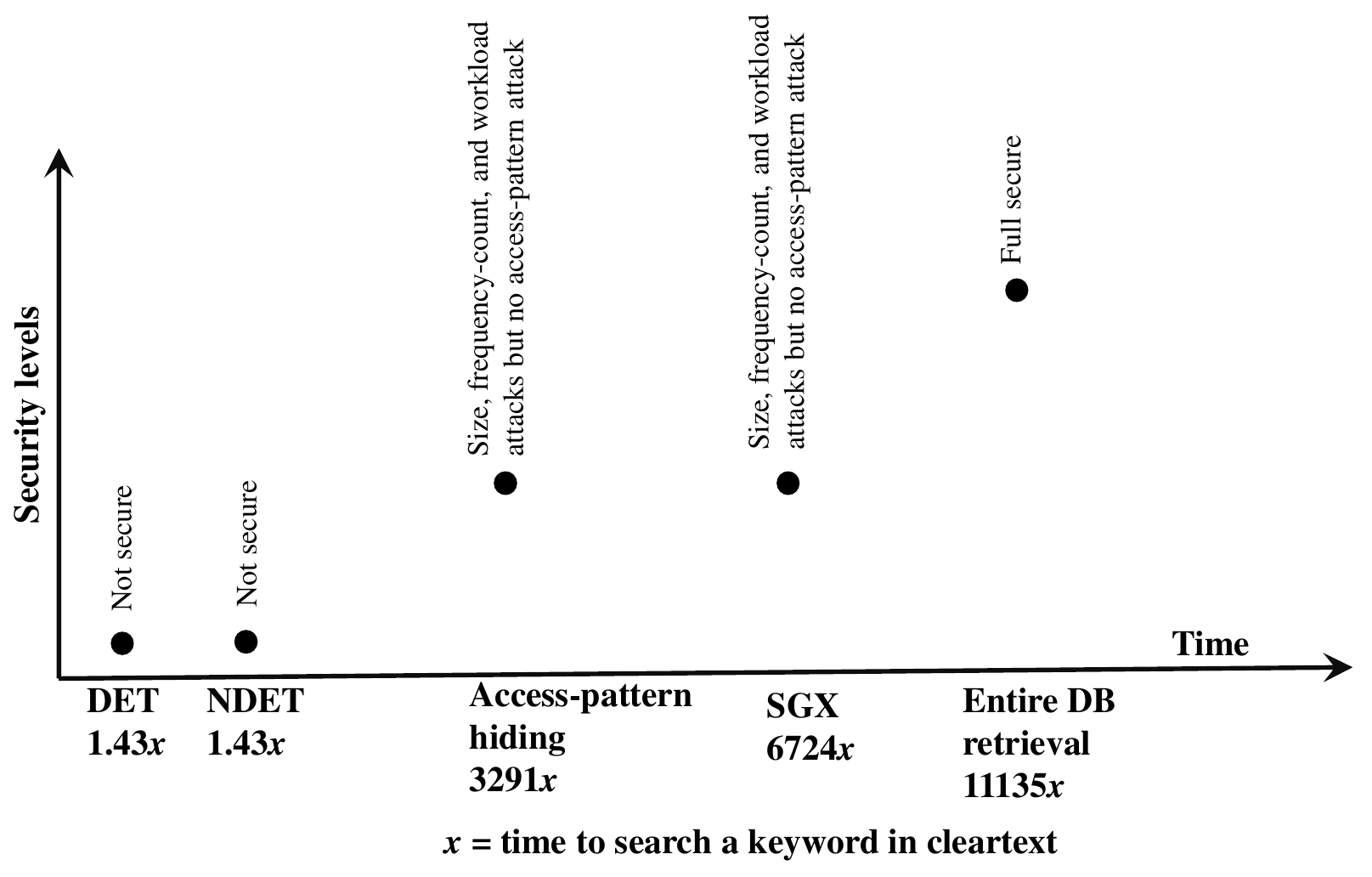}
\end{center}
The x-axis shows the ratio between the selection query execution time on encrypted data using a cryptographic technique and on cleartext data for a fixed dataset on a specific database system (in both cases), and The y-axis shows the security levels. Weak cryptographic techniques (\textit{e}.\textit{g}., deterministic encryption (DET)) are very fast but provide no security (against output size, frequency-count, access-patterns, and workload attacks), while access-pattern hiding techniques are relatively secure but slow. The completely secure technique may retrieve the entire dataset and process at the user-side but this technique is very slow. For join queries, weak cryptographic techniques are efficient since they can exploit hash/merge join. However, more secure techniques, since they need nested loop join, tends to become worse. NDET denotes non-deterministic encryption.
\end{mdframed}
\caption{Comparing different cryptographic techniques.}
\label{fig:compare}
\end{figure}

While approaches to compute over encrypted data and systems supporting such techniques are plentiful, secure data outsourcing and query processing remain an open challenge. Existing solutions suffer from several limitations. First, cryptographic approaches that prevent leakage, \textit{e}.\textit{g}., fully homomorphic encryption coupled with ORAM, simply do not scale to large data sets and complex queries for them to be of practical value. Most of the above-mentioned techniques are not developed to deal with a large amount of data and the corresponding overheads of such techniques can be very high (see Figure~\ref{fig:compare} comparing the time taken for TPC-H selection queries under different cryptographic solutions). To date, a scalable non-interactive mechanism for efficient evaluation of join queries based on homomorphic encryption that does not leak information remains an open challenge. Systems such as CryptDB~\cite{DBLP:conf/sosp/PopaRZB11} have tried to take a more practical approach by allowing users to explore the tradeoffs between the system functionality and the security it offers. Unfortunately, precisely characterizing the security offered by such systems given the underlying cryptographic approaches have turned out to be extremely difficult. For instance,~\cite{DBLP:conf/ccs/NaveedKW15,DBLP:conf/ccs/KellarisKNO16} show that when order-preserving and deterministic encryption techniques are used together, on a dataset in which the entropy of the values is not high enough, an attacker might be able to construct the entire plaintext by doing a frequency analysis of the encrypted data. While mechanisms based on secret-sharing~\cite{DBLP:journals/cacm/Shamir79} are potentially more scalable, splitting data amongst multiple non-colluding cloud operators (an assumption that is not valid in a general setting) incurs significant communication overheads and can only support a limited set of selection and aggregation queries efficiently.

While the race to develop cryptographic solutions that (\textit{i}) are efficient, (\textit{ii}) support complex SQL queries, (\textit{iii}) offer provable security from the application's perspective is ongoing, this chapter departs from the above well-trodden path by exploring a different (but complementary) approach to secure data processing by partitioning a computation over either the hybrid cloud or the public cloud based on the data classification into sensitive and non-sensitive data.
We focus on an approach for situations when only part of the data is sensitive, while the remainder (that may consist of the majority) is non-sensitive. In particular, we consider a {\bf partitioned computation model} that exploits such a classification of data into sensitive/non-sensitive subsets to develop efficient data processing solutions with {\bf provable security guarantees}. Partitioned computing potentially provides significant benefits by (\textit{i}) avoiding (expensive) cryptographic operations on non-sensitive data, and, (\textit{ii}) allowing query processing on non-sensitive data to exploit indices. 

The data classification into sensitive or non-sensitive may seem artificial/limiting at first, we refer to the readers to the ongoing dialogue in the popular media~\cite{url1,url2} about cloud security and hybrid cloud that clearly identify data classification policies to classify data as sensitive/non-sensitive as a key strategy to securing data in a cloud. Furthermore, similar to the model considered in this chapter, such articles emphasize either storing sensitive data on a private cloud while outsourcing the rest in the context of hybrid cloud or encrypting only the sensitive part of the data prior to outsourcing. Also, note that data classification based on column-level sensitivity is not a new concept. Papers~\cite{DBLP:conf/esorics/CirianiVFJPS07,DBLP:conf/esorics/CirianiVFJPS09,DBLP:journals/pvldb/VimercatiFJPS10,DBLP:journals/tissec/CirianiVFJPS10,DBLP:conf/fosad/VimercatiEFJLS13,DBLP:journals/tdsc/VimercatiFJLPS14} have explored many ways to outsource column-level partitioned data to the cloud. However, these papers does not dictate a joint query execution on two relations. Some recent database
systems such as Jana\footnote{\footnotesize \url{https://galois.com/research-development/cryptography/}} and Opaque~\cite{DBLP:conf/nsdi/ZhengDBPGS17} are exploring architectures will allow for only
some parts of the data (that is sensitive) to be encrypted while the remainder of the (non-sensitive) data remains in plaintext, thereby supporting partitioned computing. That organizational data can actually be classified as sensitive/non-sensitive is not difficult to see if we consider specific datasets. For instance, in a university dataset, data about courses, catalogs, location of classes, faculty and student enrollment would likely be not considered sensitive, but information about someone's SSN, or grade of the student would be considered sensitive.

\medskip\noindent\textbf{Contribution.} Our contributions in this chapter
are twofold:

\begin{description}

\item [Partition computation on the hybrid cloud.] Our work is motivated by recent works on the hybrid cloud that has exploited the fact that for a large class of application contexts, data can be partitioned into sensitive and non-sensitive components. Such a classification was exploited to build hybrid cloud solutions~\cite{DBLP:conf/hotcloud/KoJM11,DBLP:conf/ccs/ZhangZCWR11,DBLP:conf/ccgrid/ZhangCY14,DBLP:conf/sigmod/OktayMKK15,DBLP:conf/icde/OktayKM17} that outsource only non-sensitive data and enjoy both the benefits of the public cloud as well as strong security guarantees (without revealing sensitive data to an adversary).

\item [Partition computation on the public cloud.] In the setting of the public cloud, sensitive data is outsourced in an appropriate encrypted form, while non-sensitive data can be outsourced in cleartext form. While partitioned computing offers new opportunities for efficient and secure data processing due to avoiding cryptographic approach on the non-sensitive data, it raises several challenges when used in the public cloud. Specifically, the partitioned approach introduces a new security challenge -- that of leakage due to simultaneous execution of queries on the encrypted (sensitive) dataset and on the plaintext (non-sensitive) datasets. In this chapter, we will study such a leakage (Section~\ref{sec:Security Definition and Correctness}), a partitioned computing security definition in the context of the public cloud (Section~\ref{sec:Security Definition and Correctness}), and a way to execute partitioned data processing techniques for selection queries (Section~\ref{sec:Partitioned Computations using Existing Cryptographic Mechanisms}) that support partitioned data security while exploiting existing cryptographic mechanisms for secure processing of sensitive data and  cleartext processing of non-sensitive data. Note that the proposed approach can also be extended to other operations such as join or range queries, which are provided in~\cite{TR-2018}.

\end{description}

\section{Partitioned Computations at the Hybrid Cloud}
\label{sec:Partitioned Computations at the Hybrid Cloud}
In this section, our goal is to develop an approach to execute SQL style queries efficiently in a hybrid cloud while guaranteeing that sensitive data is not leaked to the (untrusted) public machines. At the abstract level, the technique partitions data and computation between the public and private clouds in such a way that the resulting computation (\textit{i}) minimizes the execution time, and (\textit{ii}) ensures that there is no information leakage. Information leakage, in general, could occur either directly by exposing sensitive data to the public machines, or indirectly through inferences that can be made based on selective data transferred between public and private machines during the execution.

The problem of securely executing queries in a hybrid cloud naturally leads to two interrelated subproblems:

\begin{description}

\item [Data distribution:] How is data distributed between private and public clouds? Data distribution depends on factors such as the amount of storage available on private machines, expected query workload, and whether data and query workload is largely static or dynamic.

\item [Query execution:] Given a data distribution strategy, how do we execute a query securely and efficiently across the hybrid cloud, while minimizing the execution time and obtaining the correct final outputs?
\end{description}

Since data is stored on public cloud in the clear text, data distribution strategy must guarantee that sensitive data resides only on private machines. Non-sensitive data, on the other hand, could be stored on private machines, public machines, or be replicated on both. Given a data distribution, the query processing strategy will split a computation between public and private machines while simultaneously meeting the goals  of good performance and secure execution.

\subsection{Split Strategy}
In order to ensure a secure query execution, we develop a {\em split strategy} for executing SQL queries in the hybrid cloud setting. In a split strategy, a query $Q$ is partitioned into two subqueries that can be executed {\em independently} over the private and the public cloud respectively, and the final results of the query can be computed by appropriately merging the results of the two sub-queries. In particular, a query $Q$ on dataset $D$ is split as follows:
$$	Q(D) = Q_{merge}\Big( Q_{priv}(D_{priv}), Q_{pub}(D_{pub}) \Big)$$

\noindent where $Q_{priv}$ and $Q_{pub}$ are private and public cloud sub-queries respectively. $Q_{priv}$ is executed on the private subset of $D$ (i.e., $D_{priv}$); whereas $Q_{pub}$ is performed over the public subset of $D$ (i.e., $D_{pub})$. $Q_{merge}$ is a private cloud merge sub-query that reads the outputs of former two sub-queries as input and creates the outputs equivalent to that of original $Q$. We call such an execution strategy as \textit{split-strategy}.

Two aspects of \textit{split-strategy} are noteworthy:
\begin{enumerate}
 \item
It offers full security, since the public machines only have access to $D_{pub}$ that do not contain any sensitive data. Moreover, no information is exchanged between private and public clouds during the execution of $Q_{pub}$, resulting in the execution at the public cloud to be \textit{observationally equivalent} to the situation where $D_{priv}$ could be any random data.

\item
Split-strategy gains efficiency by executing $Q_{priv}$ and $Q_{pub}$ in parallel at the private and public cloud respectively, and furthermore, by performing inter-cloud data transfer at most once throughout the query execution.
Note that the networks between private and public clouds can be significantly slower compared to the networks used within clouds. Thus, minimizing the amount of data shuffling between the clouds will have a big performance impact.
\end{enumerate}
Split strategy, and its efficiency, depends upon the data distribution strategy used to partition the data between private and public clouds. Besides storing sensitive data, the private cloud must also store part of non-sensitive data (called \emph{pseudo sensitive data}) that may be needed on the private side to support efficient query processing. For instance, a join query may necessitate that non-sensitive data be available at the private node in case-sensitive records from one relation may join with non-sensitive records in another. Since in the split-execution strategy, the two subqueries execute independently with no communication, if we do not store non-sensitive data at the private side, we will need to transfer entire relation to the private side for the join to be computed as part of the merge query.

\smallskip\noindent\textbf{Split-strategy for selection or projection.} An efficient \textit{split-strategy} for selection or projection operation is straightforward. In this case, $Q_{priv}$ is equivalent to the original query $Q$, but is performed only over sensitive records in $D_{priv}$. Likewise, $Q_{pub} = Q$, but only runs over $D_{pub}$. Finally, $Q_{merge} = Q_{priv} \cup Q_{pub}$.

\begin{figure}[h]
\centering
\includegraphics[scale=0.45]{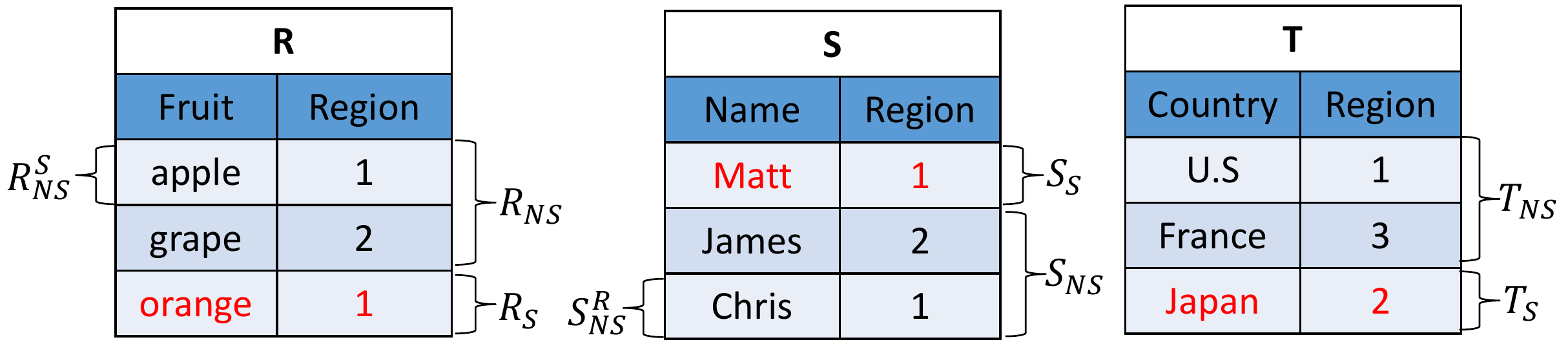}
\hfill
\caption{Example relations.}
\label{fig:examplerelations}
\end{figure}

\smallskip\noindent\textbf{Split-strategy for equijoin.} An efficient \textit{split-strategy} for performing a join query such as $Q= R \underset{C}{\bowtie}S$ is more complex. To see this, consider the relations $R$ and $S$ as shown above in Figure~\ref{fig:examplerelations}, where sensitive portions of $R$ and $S$ are denoted as $R_s$ and $S_s$, respectively, and remaining fraction of them are non-sensitive, denoted as $R_{ns}$ and $S_{ns}$, and the join condition is $C = (R.region = S.Region)$. Let us further assume that
$R_{ns}$ and $S_{ns}$, besides being stored in the public cloud are also replicated on the private cloud.

The \textit{naive split-strategy} for $R \underset{C}{\bowtie}S$ would be:
\begin{itemize}
\item $Q_{pub} = R_{ns} \underset{C}{\bowtie} S_{ns}$
\item $Q_{priv} = (R_{s} \underset{C}{\bowtie} S_{s}) \cup (R_{s} \underset{C}{\bowtie} S_{ns}) \cup (R_{ns} \underset{C}{\bowtie} S_s$).
\end{itemize}

Note that if $Q$ is split as above, $Q_{priv}$ consists of three subqueries which scan 2, 3, and 3 tuples in $R$ and $S$ respectively resulting in 8 tuples to be scanned and joined. In contrast, if we simply executed the query $Q$ on the private side (notice that we can  do so, since $R$ and $S$ are fully stored on the private side), it would result in lower cost requiring scan of 6 tuples on the private side. Indeed, the overhead of the above split strategy increases even further if we consider multiway joins (e.g., $R \underset{C}{\bowtie} S \underset{C^{\prime}}{\bowtie} T$) compared to simply implementing the multiway join locally. Thus, if we use split-strategy for  computing $R \underset{C}{\bowtie} S \underset{C^{\prime}}{\bowtie} T$, where $C^{\prime}$ is $S.Region=T.Region$, then the number of tuples that are scanned/joined in the private cloud will be much higher than that of the original query.

\smallskip\noindent\textbf{A modified approach for equijoin.} The cost of executing $Q$ in the private cloud can be significantly reduced by pre-filtering  relations $R$ and $S$ based on sensitive records of the other relation. To perform such a pre-filtering operation, the tuples in the relations $R_{ns}$ and $S_{ns}$ have to be co-partitioned based on whether they join with a sensitive tuple from the other table under condition $C$ or not.

Let $R^{S}_{ns}$ be a set of non-sensitive tuples of $R$ that join with any sensitive tuple in $S$. In our case, $R^{S}_{ns} = \langle$apple, 1$\rangle$. Similarly, let $S^{R}_{ns}$ be non-sensitive tuples of $S$ that join with any record from $R_s$, \textit{i}.\textit{e}., $\langle$Chris, 1$\rangle$. In that case, the new private side computation can be rewritten as:
\begin{equation}
\label{eq:nofiltering}
(R_s \cup R^{S}_{ns}) \underset{C}{\bowtie} (S_s \cup S^{R}_{ns}).
\end{equation}
Thus, the scan and join cost of this new plan at the private cloud is 4, which is lower compared to computing the query entirely on the private side that had a cost of 6.


\smallskip\noindent\textbf{Guarded join.} The above mentioned modified strategy, nonetheless, introduces a new challenge. Since $R^{S}_{ns} \underset{C}{\bowtie} S^{R}_{ns}$ is both repeated at public and private cloud, the output of $R^{S}_{ns} \underset{C}{\bowtie} S^{R}_{ns}$, $\langle$apple, Chris, 1$\rangle$, is computed on both private and public clouds. To prevent this, we do a guarded join ($\bowtie^{\prime}$) on the private cloud, which discards the output, if it is generated via joining two non-sensitive tuples. This feature can easily be implemented by adding a column to $R$ and $S$ that marks the sensitivity status of a tuple, whether it is sensitive or non-sensitive, and then by adding an appropriate selection after the join operation. In other words, the complete representation of private side computation for $R \underset{C}{\bowtie} S$ would be
\begin{equation}
\sigma_{R.sens = true \vee S.sens = true}((R_s \cup R^{S}_{ns}) \underset{C}{\bowtie} (S_s \cup S^{R}_{ns}))
\end{equation}

\noindent where $sens$ is a boolean column (or partition id) appended to relations $R$ and $S$ on the private cloud. Assume that it is set to true for sensitive records and false for non-sensitive records.

\smallskip\noindent\textbf{Challenges.} There exist multiple  challenges in implementing this new approach. First challenge is the cost of creating $R^{S}_{ns}$ and $S^{R}_{ns}$ beforehand. Extracting these partitions for a query might take as much time as executing the original query. However, the costs are amortized since these relations are computed once, and used multiple times to improve join performance at the private cloud.

The second challenge is the creation of co-partitioning tables for complex queries. For instance, in case of a query $R \underset{C}{\bowtie} S \underset{C^{\prime}}{\bowtie} T$, the plan would be to first compute results of $R \underset{C}{\bowtie} S$, and then to join them with $T$. However, if we do the private side computation of $R \underset{C}{\bowtie} S$, based on Equation~\ref{eq:nofiltering} (no duplicate filtering) and join the results with $T$, then we will not be able to obtain the complete set of sensitive $R \underset{C}{\bowtie} S \underset{C^{\prime}}{\bowtie} T$ results.

To see this, consider the sensitive record (in Figure~\ref{fig:examplerelations}) $\langle$Japan, 2$\rangle$ in $T$ that joins with non-sensitive $\langle$grape, 2$\rangle$ tuple in $R- R^{S}_{ns}$ or joins with non-sensitive $\langle$James, 2$\rangle$ tuple from $S-S^{R}_{ns}$. Thus, the non-sensitive records of $R$ and $S$ has to be co-partitioned based on the sensitive records of $T$ via their join paths from $T$. In $R\underset{C}{\bowtie}S \underset{C^{\prime}}{\bowtie} T$, the join path from $T$ to $R$ is $T \underset{C^{\prime}}{\bowtie} S \underset{C}{\bowtie} R$ and from $T$ to $S$ is $T \underset{C^{\prime}}{\bowtie} S$. Similarly, the non-sensitive $T$ records has to be co-partitioned based on the sensitive $R$ and $S$ records via join paths specified in the query.

Final challenge is in maintaining these co-partitions and feeding the right one when an arbitrary query arrives. Given a workload of queries and multiple possible join paths between any two relations, each relation $R$ in the dataset may need to be co-partitioned multiple times. This implies that any non-sensitive record $r$ of $R$ might appear in more than one co-partition of $R$. So, maintaining each co-partition separately might be unfeasible in terms of storage. However, the identifiers of each co-partition that record $r$ belongs to can be embedded into $r$ as a new column. We call such a column as the \textit{co-partition} (CPT) column. Note that CPT column is \textit{only defined} on the private cloud data, since revealing it to public cloud would violate our security requirement.

CPT column initially will be set to null for sensitive tuples in the private side, since the co-partitions are only for non-sensitive tuples. Thus, it can further be used to serve another purpose, indicating the sensitivity status of a tuple $r$ by setting it to ``sens" only for sensitive tuples.

\smallskip\noindent\textbf{Join path.} To formalize the concept of co-partitioning, we first need to define the notion of join path. Let $R_{i}$ be a relation in our dataset $D$, and let $Q$ be a query over the relation $R_i$. We say a join path exists from a relation $R_j$ to $R_i$, if either $R_{i}$ is joined with $R_j$ directly based on a condition $C$, \textit{i}.\textit{e}., $R_j \underset{C}{\bowtie} R_i$, or $R_{j}$ is joined with $R_i$ \textit{indirectly} using other relations in $Q$. A join path $p$ can be represented as a sequence of relations and conditions between $R_{j}$ and $R_{i}$ relations. Let $PathSet$ be the set of all join paths that are extracted either from the expected workload or a given dataset schema.
\begin{equation}
PathSet_{i} = \{ \forall p \in PathSet : \text{path } p \text{ ends at relation } R_i \}.
\end{equation}

Let $CP(R_{i}, p)$ be the set of non-sensitive $R_{i}$ records that will be joined with at least one sensitive record from any other relation $R_j$ via the join path $p$. Note that $p$ starts from $R_j$ and ends at $R_i$ that can be used as an id to $CP(R_{i}, p)$. Any $CP(R_{i}, p)$ is called as ``co-partition'' of $R_{i}$. Given these definitions, the CPT column of a $R_{i}$ record, say $r$, can be defined as:
\begin{equation}
r.CPT = \begin{cases}
	sens & \text{if } r \text{ is sens.} \\
	\{ \forall p \in PathSet_{i}: r \in CP(R_{i}, p)\} & \text{otherwise}
	\end{cases}
\end{equation}
\normalsize

\begin{figure}[h]
\centering
\includegraphics[scale=0.4]{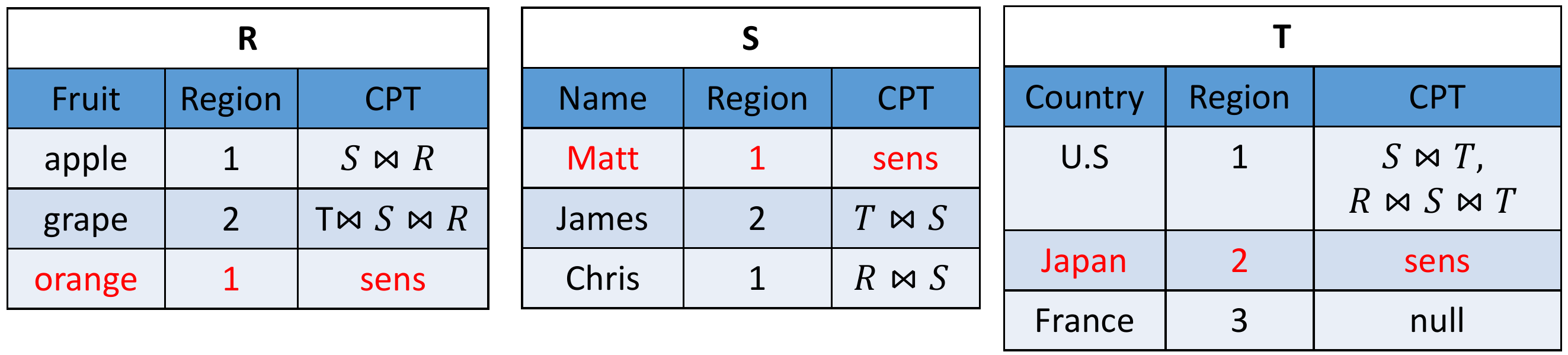}
\caption{Example relations with the CPT columns.}
\label{fig:examplerelationswithCPTColumn}
\BB
\BB
\end{figure}

Figure~\ref{fig:examplerelationswithCPTColumn} shows our example $R$, $S$ and $T$ relations with their CPT column. For instance, the join path $R \bowtie S$ will be appended to the CPT column of all the tuples in $ S^{R}_{ns}$. Additionally, the CPT column of all tuples in $R_s$ will be set to $sens$.

\subsection{Experimental Analysis}

To study the impact of table partitioning discussed in the previous section, we differentiate between two realizations of our strategy: in our first technique, entitled (\textit{CPT-C}), every record in a table at the private cloud contains a CPT column and they are physically stored together; whereas in our second approach, entitled \textit{CPT-P}, the tables are partitioned based on their record's CPT column and each partition is stored separately. Each partition file then appended to the corresponding Hive table as a separate partition, so at querying stage, Hive filters out the unnecessary partitions for that particular query.

\begin{figure}[!h]
    \begin{minipage}[t]{0.45\linewidth}
    \centering
    \includegraphics[scale=0.4]{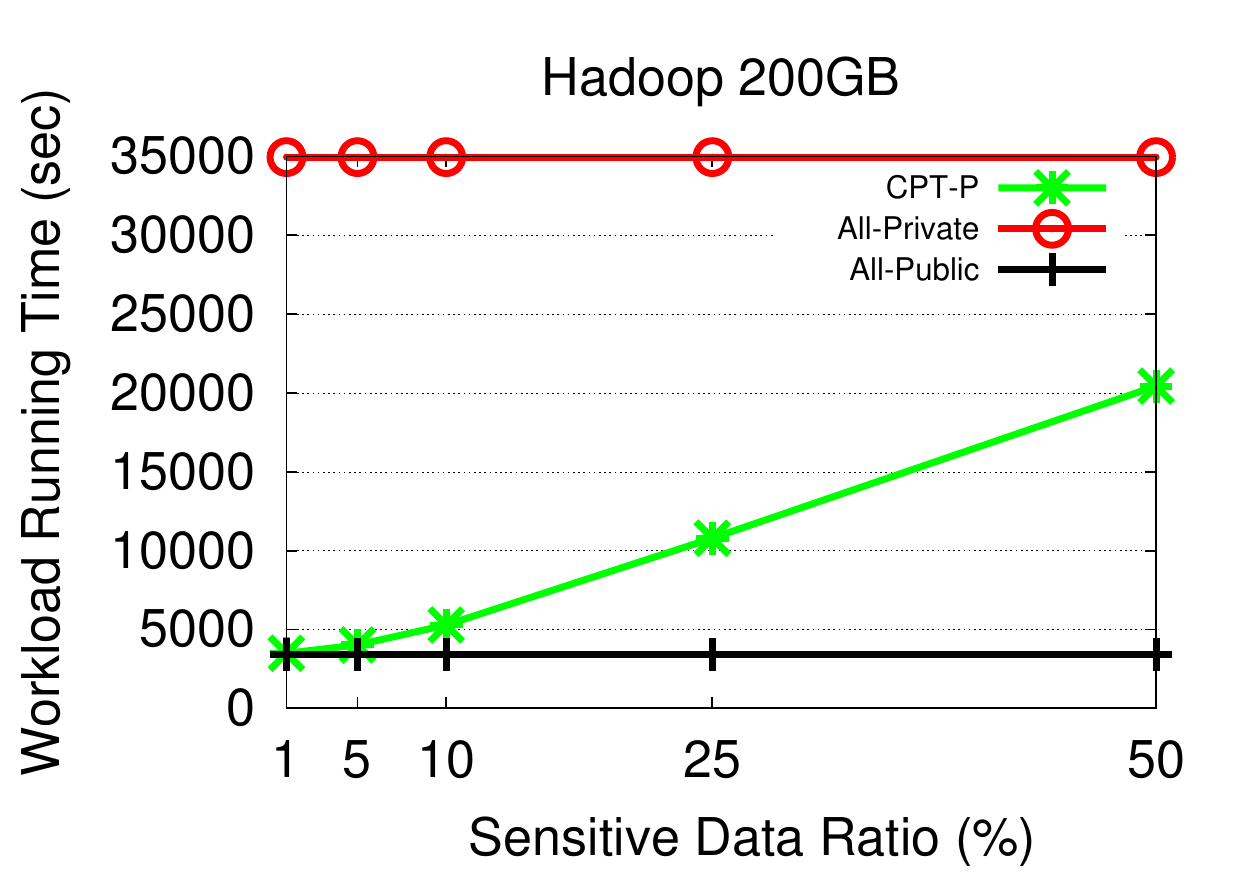}
    \end{minipage}
\quad\quad
    \begin{minipage}[t]{0.49\linewidth}
    \centering
    \includegraphics[scale=0.4]{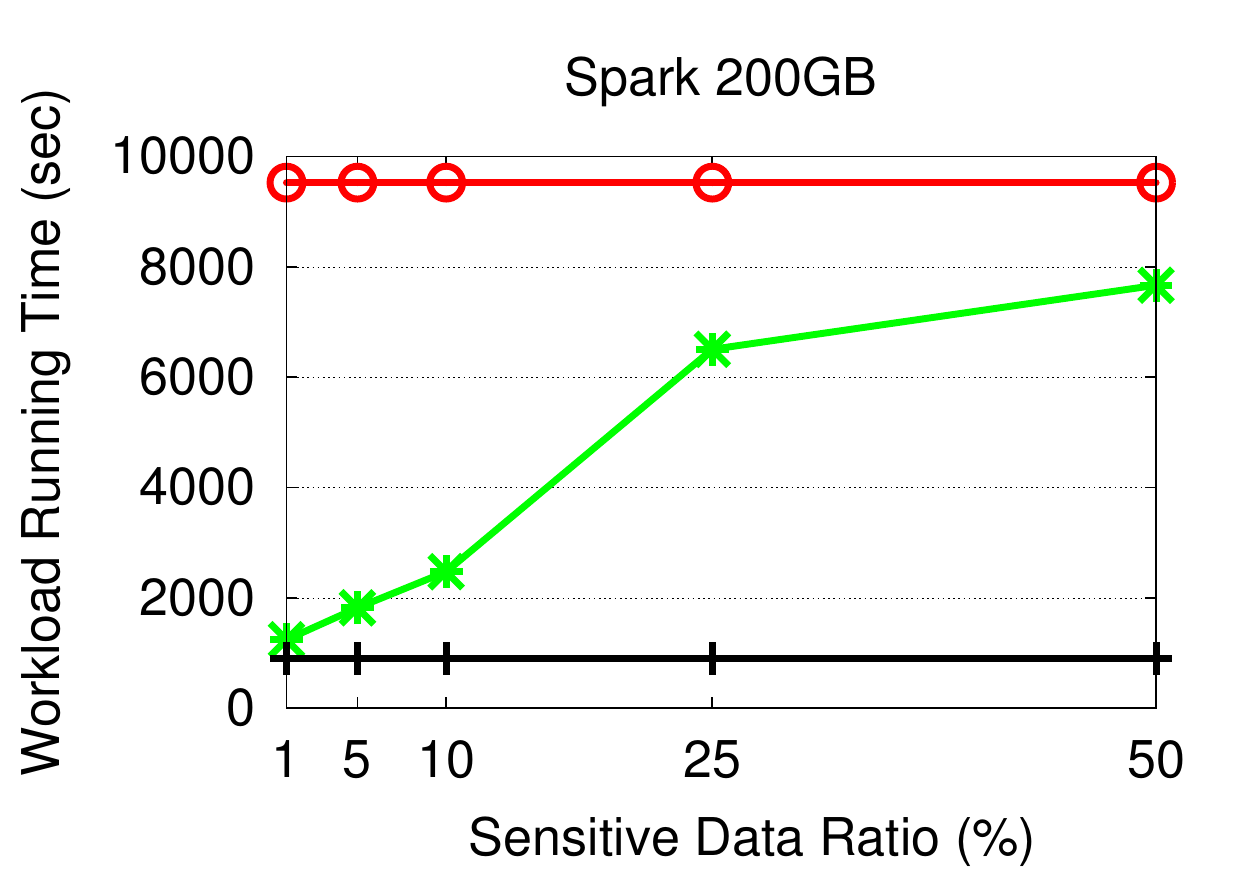}
    \end{minipage}
\caption{Running times for different sensitivity ratios.}
\label{fig:DifferentSensitivityRatios}
\end{figure}

\smallskip\noindent\textbf{Sensitive data ratio.} For these experiments, we varied the amount of sensitive records ($1,5,10,25,50\%$) in \textit{customer} and \textit{supplier} tables. Also, we set the number of public machines to $36$. As expected, Figure~\ref{fig:DifferentSensitivityRatios} shows that a larger percentage of sensitive data within the input leads to a longer workload execution time for both, CPT-C and CPT-P in Hadoop and Spark. The reason behind this is that a higher sensitive data ratio results in more computations being performed on the private side and implies a longer query execution time in \textit{split-strategy}. When the sensitivity ratio increases, CPT-P's scan cost increases dramatically. Since the scan cost of queries is the dominant factor compared to other operators (join, filtering etc.) in Spark, CPT-C provides a very low-performance gain compared to All-Private in Spark. Because the scan cost of these two approaches is same. Overall, when sensitivity ratio is as low as $1\%$, CPT-P provides $8.7\times$ speed-up in Hadoop and $5\times$ speed-up in Spark compared to All-Private. 

\begin{figure}[h]
    \begin{minipage}[t]{0.45\linewidth}
    \centering
    \includegraphics[scale=0.4]{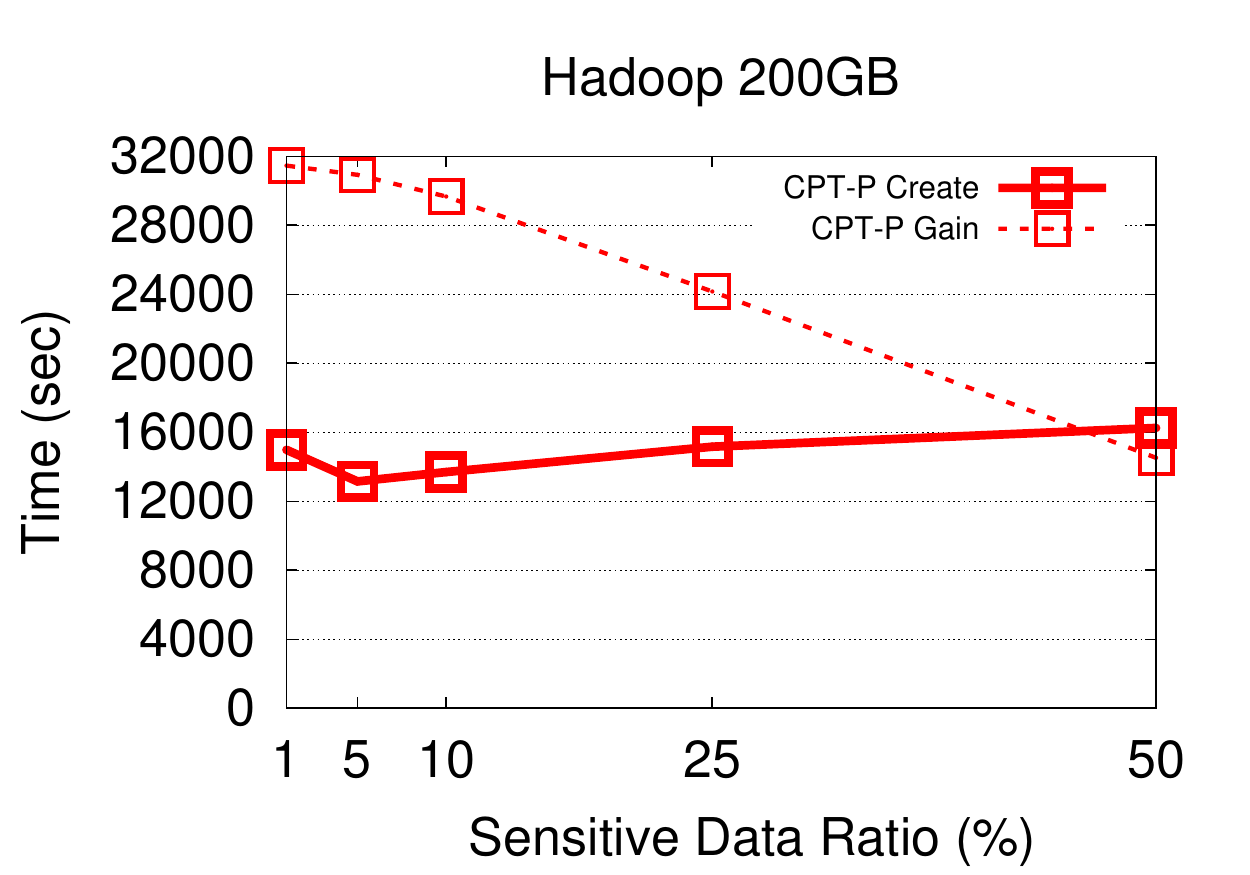}
    \end{minipage}
\quad\quad
    \begin{minipage}[t]{0.49\linewidth}
    \centering
    \includegraphics[scale=0.4]{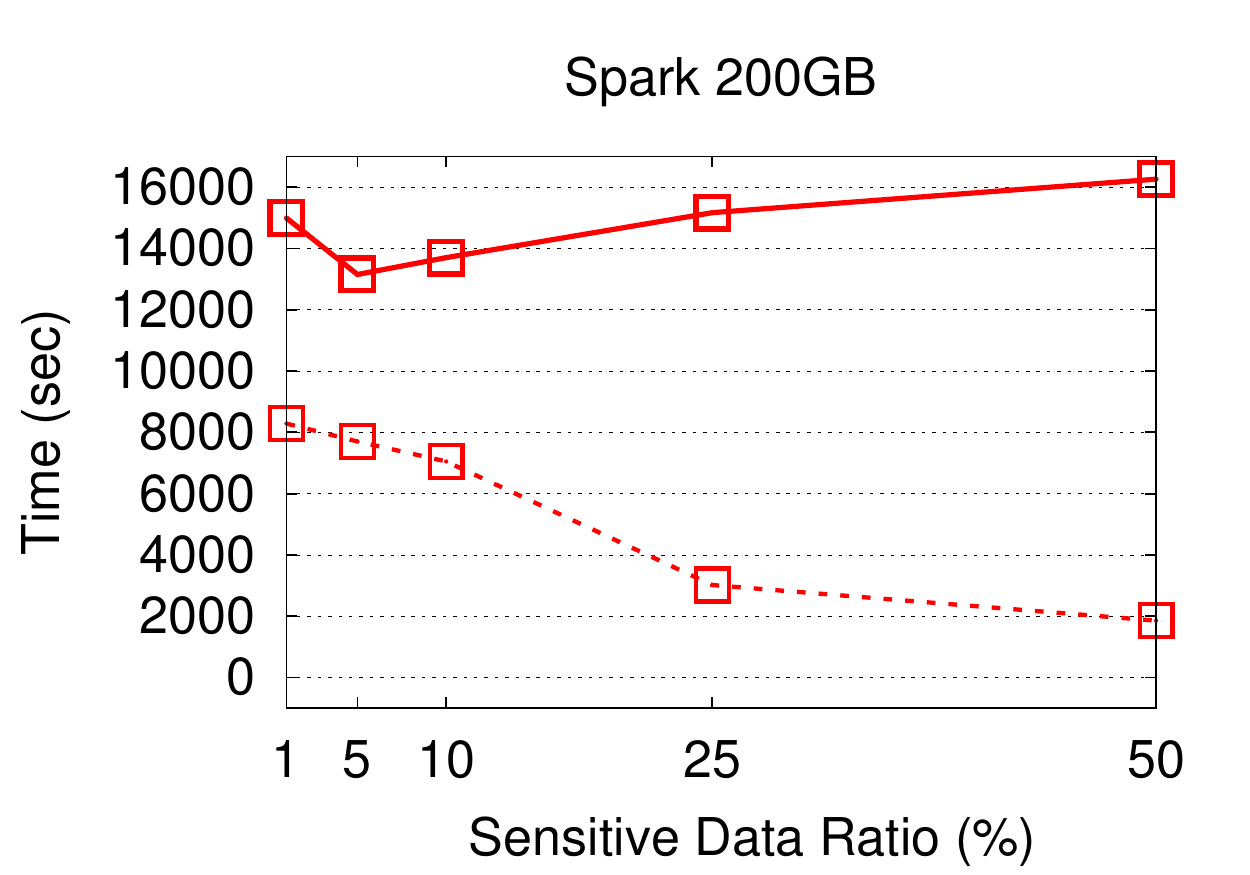}
    \end{minipage}
\caption{The CPT column's creation for different sensitivity ratios.}
\BB\label{fig:CPTCreation}
\end{figure}

\begin{figure}[!h]
    \centering
    \includegraphics[scale=0.5]{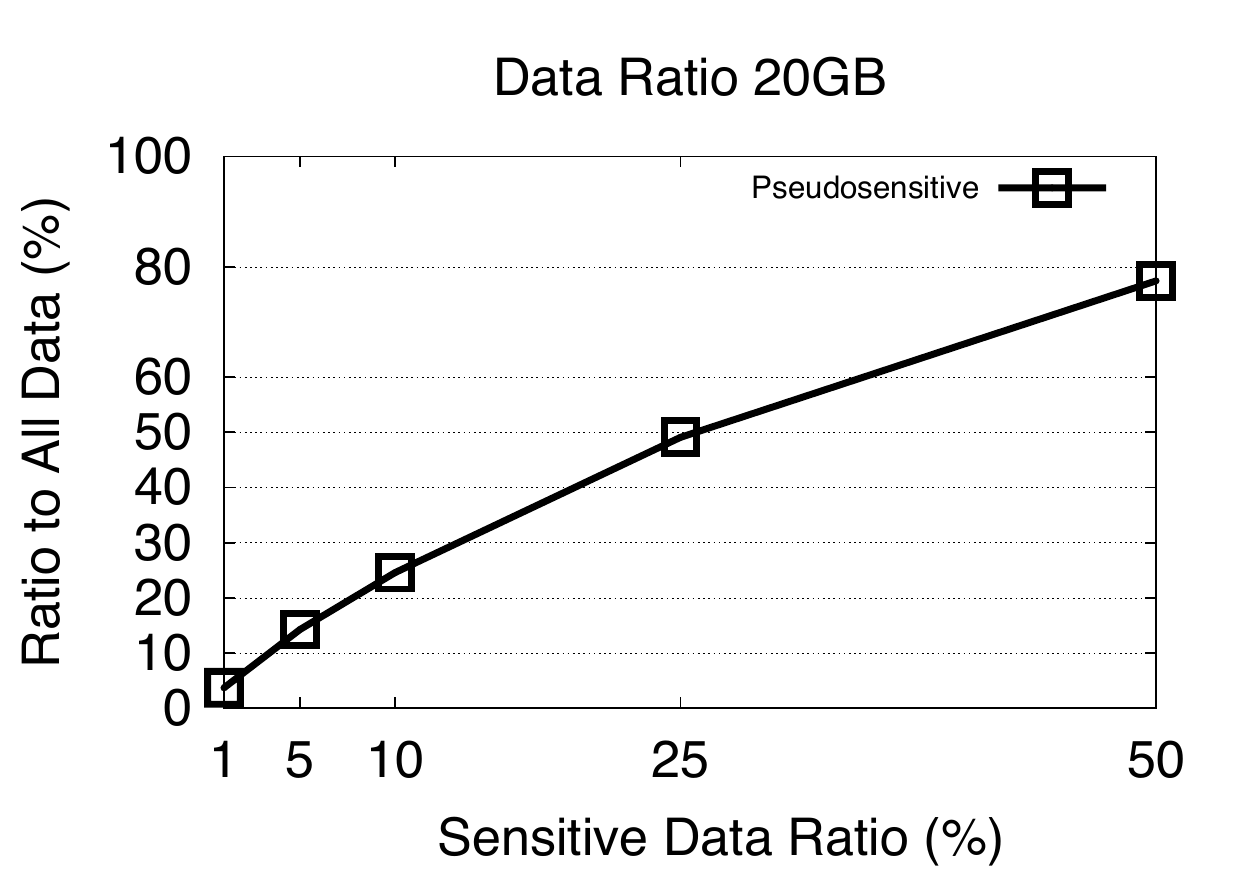}
    \caption{Comparison of pseudo-sensitive data and sensitivity ratio.}
\label{fig:pseudo-s}
\end{figure}

Recall that we created the CPT column using a Spark job for CPT-C solution. We then physically partitioned tables for CPT-P solution. Figure~\ref{fig:CPTCreation} shows how much time we spent in preparing private cloud data for both CPT-C and CPT-P. It also indicates the gains of these approaches compared to All-Private in terms of the overall workload execution time. As indicated in Figure~\ref{fig:CPTCreation}, until $25\%$ sensitivity, CPT-P's data preparation time is less than that of performance gain in Hadoop; whereas in Spark, data preparation time is always higher than the performance gain for both CPT-P and CPT-C. 
Note that, we prepare the CPT column only once on a static data for an expected workload that will more likely be executed more than once with different selection and projection conditions. In Spark, if the sensitivity ratio is as high as $10\%$, executing the workload more than once will be enough for the performance gain of CPT-P solution to be higher than the overhead of data preparation time.

\smallskip
\noindent {\bf Size of Private Storage.}
Besides storing sensitive data, in our technique, we also store pseudo-sensitive data on the private cloud.  This enables us to execute queries in a partitioned manner  while minimizing expensive inter-cloud communication during query execution. In Figure~\ref{fig:pseudo-s}, we plot the size of pseudo-sensitive data as a percentage of total database size at different sensitivity levels. We note that even when sensitivity levels are as high as 5-10\%, the pseudo-sensitive data remains only a fraction (15-25\% of the total data). At smaller sensitivity levels, the ratio is much smaller.

\subsection{Other Approaches to Partitioned Computing}
The discussion above focused on partitioned computing in hybrid clouds in the context of SQL queries and is based primarily on the work that appeared in \cite{DBLP:conf/icde/OktayKM17}. Several other approaches to partitioned computing in the hybrid cloud have also been developed in the literature that, similar to the above-mentioned method, offer security by controlling data distribution between private and public clouds. Many of these approaches~\cite{DBLP:conf/hotcloud/KoJM11,DBLP:conf/ccs/ZhangZCWR11,DBLP:conf/ccgrid/ZhangCY14,DBLP:conf/sigmod/OktayMKK15}  have been developed in the context of MapReduce job execution, and they address security at a lower level compared to the approach defined above, which is at SQL level. Note that one could, potentially, transform SQL/Hive queries into lower level MapReduce jobs and run such MapReduce jobs using privacy preserving extensions. There are several limitations of such an approach, however, and we refer the reader to \cite{DBLP:conf/icde/OktayKM17} for a detailed discussion of the limitations of such an approach and to~\cite{DBLP:journals/csr/DerbekoDGS16} for a detailed survey on the hybrid cloud based MapReduce security.

\section{Partitioned Computations at the Public Cloud and Security Definition}
\label{sec:Security Definition and Correctness}
In this section, we define the partitioned computation, illustrate how such a computation can leak information due to the joint processing of sensitive and non-sensitive data, discuss the corresponding security definition, and finally discuss system and adversarial models under which we will develop our solutions.

\subsection*{Partitioned Computations}
\label{subsec:Partitioned Computations}
Let $R$ be a relation that is partitioned into two sub-relations, $R_e\supseteq R_s$ and $R_p \subseteq R_{\mathit{ns}}$, such that $R= R_e \cup R_p$. The relation $R_e$ contains all the sensitive tuples (denoted by $R_s$) of the relation $R$ and will be stored in encrypted form in the cloud. Note that $R_e$ may contain additional (non-sensitive) tuples of $R$, if that helps with secure data processing). The relation $R_p$ refer to the sub-relation of $R$ that will be stored in plaintext on the cloud. Naturally, $R_p$ does not contain any sensitive tuples. For the remainder of the chapter, we will assume that $R_e=R_s$ and $R_p=R_{ns}$, though our approach will be generalized to allow for a potentially replicated representation of non-sensitive data in encrypted form, if it helps to evaluate queries more efficiently. Let us consider a query $Q$ over relation $R$. A partition computation strategy splits the execution of $Q$ into two independent sub-queries: $Q_s$: a query to be executed on $E(R_e)$ and $Q_{\mathit{ns}}$: a query to be executed on $R_{p}$. The final results are computed (using a query $Q_{merge}$) by appropriately merging the results of the two sub-queries at the trusted database (DB) owner side (or in the cloud, if a trusted component, \textit{e}.\textit{g}., Intel SGX, is available for such a merge operation). In particular, the query $Q$ on a relation $R$ is partitioned, as follows:
$$Q(R) = Q_{merge}\Big( Q_s(R_e), Q_{\mathit{ns}}(R_{\mathit{p}}) \Big)$$
\noindent Let us illustrate partitioned computations through an example.

\begin{figure}[h]
\centering
\begin{tabular}{|l||l|l|l|l|l|l|l|}
    \hline
      &EId & FirstName & LastName & SSN  & Office\# & Department \\ \hline
    $t_1$& E101 & Adam & Smith       & 111&1 & Defense  \\ \hline
    $t_2$& E259 & John & Williams   & 222&2 & Design   \\ \hline
    $t_3$& E199 & Eve  & Smith      & 333&2  & Design   \\ \hline
    $t_4$& E259 & John & Williams   & 222&6 & Defense   \\ \hline
    $t_5$& E152 & Clark & Cook   & 444&1 & Defense   \\ \hline
    $t_6$& E254 & David & Watts   & 555&4 & Design   \\ \hline
    $t_7$& E159 & Lisa & Ross   & 666&2 & Defense   \\ \hline
    $t_8$& E152 & Clark & Cook   & 444&3 & Design   \\ \hline
  \end{tabular}
    \caption{A relation: \textit{Employee}.}
  \label{fig:employee relation}
\end{figure}

\textnormal{\textbf{Example 1}}: Consider an \emph{Employee} relation, see Figure~\ref{fig:employee relation}. In this relation, the attribute {\em SSN} is sensitive, and furthermore, all tuples of employees for the {\em Department} $=$ ``\texttt{Defense}'' are sensitive. In such a case, the \emph{Employee} relation may be stored as the following three relations: (\textit{i}) \emph{Employee1} with attributes {\em EId} and {\em SSN} (see Figure~\ref{fig:employee1 relation}); (\textit{ii}) \emph{Employee2} with attributes {\em EId}, {\em FirstName}, {\em LastName}, {\em Office\#}, and {\em Department}, where {\em Department} $=$ ``\texttt{Defense}'' (see Figure~\ref{fig:employee2 relation}); and (\textit{iii}) \emph{Employee3} with attributes {\em EId}, {\em FirstName}, {\em LastName}, {\em Office\#}, and {\em Department}, where {\em Department} $<>$ ``\texttt{Defense}'' (see Figure~\ref{fig:employee3 relation}). Since the relations \emph{Employee1} and \emph{Employee2} (Figures~\ref{fig:employee1 relation} and~\ref{fig:employee2 relation}) contain only sensitive data, these two relations are encrypted before outsourcing, while \emph{Employee3} (Figure~\ref{fig:employee3 relation}), which contains only non-sensitive data, is outsourced in clear-text. We assume that the sensitive data is strongly encrypted such that the property of \emph{ciphertext indistinguishability} (\textit{i}.\textit{e}., an adversary cannot distinguish pairs of ciphertexts) is achieved. Thus, the two occurrences of \texttt{E152} have two different ciphertexts.

\begin{figure}[!h]
  \centering
\begin{tabular}{|l||l|l|}\hline
         & EId  & SSN   \\\hline
    $t_1$& E101 & 111   \\\hline
    $t_2$& E259 & 222   \\\hline
    $t_3$& E199 & 333   \\\hline
    $t_5$& E152 & 444   \\\hline
    $t_6$& E254 & 555   \\\hline
    $t_7$& E159 & 666   \\\hline
  \end{tabular}
  \caption{A sensitive relation: \textit{Employee1}.}
    \label{fig:employee1 relation}
  \end{figure}

\begin{figure}[!h]
  \centering
\begin{tabular}{|l||l|l|l|l|l|l|}\hline
         & EId  & FirstName & LastName &  Office\# & Department \\ \hline
    $t_1$& E101 & Adam & Smith       & 1 & Defense  \\ \hline
    $t_4$& E259 & John & Williams   & 6 & Defense   \\ \hline
    $t_5$& E152 & Clark & Cook   &1 & Defense   \\ \hline
    $t_7$& E159 & Lisa & Ross   &2 & Defense   \\ \hline
  \end{tabular}
  \caption{A sensitive relation: \textit{Employee2}.}
\label{fig:employee2 relation}
\end{figure}

\begin{figure}[!h]
\centering
\begin{tabular}{|l||l|l|l|l|l|l|}
    \hline
         & EId  & FirstName & LastName &  Office\# & Department \\ \hline
    $t_2$& E259 & John & Williams   &2 & Design   \\ \hline
    $t_3$& E199 & Eve  & Smith      &2  & Design   \\ \hline
    $t_6$& E254 & David & Watts   & 4 & Design   \\ \hline
    $t_8$& E152 & Clark & Cook   & 3 & Design   \\ \hline
  \end{tabular}
  \caption{A non-sensitive relation: \textit{Employee3}.}
\label{fig:employee3 relation}
  \end{figure}

Consider a query \texttt{Q: SELECT FirstName, LastName, Office\#, Department from Employee where FirstName = ''John''}. In partitioned computation, the query \texttt{Q} is partitioned into two sub-queries: $Q_s$ that executes on \texttt{Employee2}, and $Q_{ns}$ that executes on \texttt{Employee3}. $Q_s$ will retrieve the tuple $t_4$ while $Q_{\mathit{ns}}$ will retrieve the tuple $t_2$. $Q_{merge}$ in this example is simply a union operator. Note that the execution of the query \texttt{Q} will also retrieve the same tuples.

\subsection*{Inference Attack in Partitioned Computations}
\label{subsec:Inference Attack and Partitioned Data Security}
Partitioned computations, if performed naively, could lead to inferences about sensitive data from non-sensitive data. To see this, consider following three queries on the \emph{Employee2} and \emph{Employee3} relations: (\textit{i}) retrieve tuples of the employee \texttt{Eid = E259}, (\textit{ii}) retrieve tuples of the employee \texttt{Eid = E101}, and (\textit{iii}) retrieve tuples of the employee \texttt{Eid = E199}. We consider an \emph{honest-but-curious} adversarial cloud that returns the correct answers to the queries but wishes to know information about the encrypted sensitive tables, \emph{Employee1} and \emph{Employee2}.

Table~\ref{tab:answer table}
shows the adversary's view based on executing the corresponding $Q_s$ and $Q_{ns}$ components of the above three queries assuming that the tuple retrieving cryptographic approaches are not hiding access-patterns. During the execution, the adversary gains complete knowledge of non-sensitive tuples returned, and furthermore, knowledge about which encrypted tuples are returned as a result of $Q_s$ ($\mathit{E(t_i)}$ in the table refers to the encrypted tuple $t_i$).

\begin{table}
\centering
    \begin{tabular}{|l|l|l|}
    \hline
    Query value & \multicolumn{2}{|c|}{Returned tuples/Adversarial view}           \\ \hline
    ~                          & Employee2 & Employee3 \\ \hline
    E259                      & $\mathit{E(t_4)}$        & $t_2$        \\ \hline
    E101                              & $\mathit{E(t_1)}$        & null      \\ \hline
    E199                            & null        & $t_3$      \\ \hline
    \end{tabular}
    \caption{Queries and returned tuples/adversarial view.}
    \label{tab:answer table}
\end{table}

Given the above adversarial view, the adversary learns that employee \texttt{E259} has tuples in
both $D_s$ ($= D_e$) and $D_p$ ($=D_{ns}$). Coupled with the knowledge about data partitioning, the adversary can learn that \texttt{E259} works in both sensitive and non-sensitive departments. Moreover, the adversary learns which sensitive tuple has an \emph{Eid} equals to \texttt{E259}.
From the 2nd query, the adversary learns that \texttt{E101} works only in a sensitive department, (since the query did not return any answer from the Employee3 relation). Likewise, from the 3rd query, the adversary learns that \texttt{E199} works only in a non-sensitive department.

\begin{figure}[!h]
\centering
\includegraphics[scale=0.7]{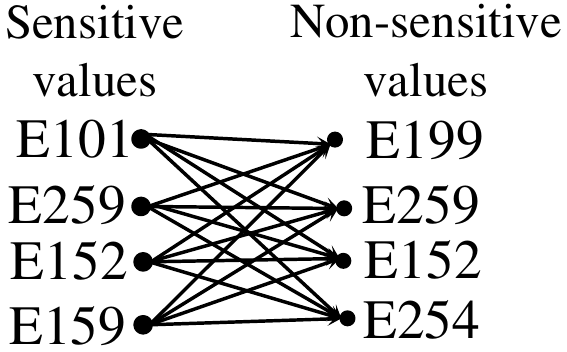}
  \caption{A bipartite graph showing an initial condition sensitive and non-sensitive values before query execution.}
  \label{fig:survival_matching1}
\end{figure}

In order to prevent such an attack, we need a new security definition. Before we discuss the formal definition of partitioned data security, we first provide intuition for the definition. Observe that before retrieving any tuple, under the assumption that no one except the DB owner can decrypt an encrypted sensitive value, say $E(s_i)$, the adversary cannot learn which non-sensitive value is identical to cleartext value of $E(s_i)$; let us denote $s_i$ as cleartext of $E(s_i)$. Thus, the adversary will consider that the value $s_i$ is identical to one of the non-sensitive values. Based on this fact, the adversary can create a complete bipartite graph having $|S|$ nodes on one side and $|\mathit{NS}|$ nodes on the other side, where $|S|$ and $|\mathit{NS}|$ are a number of sensitive and non-sensitive values, respectively. The edges in the graph are called \emph{surviving matches of the values}. For example, before executing any query, the adversary can create a bipartite graph for 4 sensitive and 4 non-sensitive values of \texttt{EID} attribute of Example 1; as shown in Figure~\ref{fig:survival_matching1}.

The query execution on the datasets creates an adversarial view that guides the adversary to create a (new) bipartite graph of the same number of nodes on both sides. The requirement is to preserve all the edges of the initial bipartite graph in the graph obtained after the query execution, leading to the initial condition that the cleartext of the value $E(s_i)$ is identical to one of the non-sensitive values. Note that if the query execution removes any surviving matches of the values, it will leak that the value $s_i$ is not identical to those non-sensitive values.

We also need to hide occurrences of a sensitive value. Before a query execution, due to ciphertext indistinguishability, all occurrences of a single sensitive value are different, but a simple search or join query may reveal how many tuples have the same value. Based on the above two requirements, we can define a notion of \emph{partitioned data security}.

\subsection*{Partitioned Data Security at the Public Cloud}
Let $R$ be a relation containing sensitive and non-sensitive tuples. Let $R_s$ and $R_{\mathit{ns}}$ be the sensitive and non-sensitive relations, respectively. Let $q(R_s,R_{\mathit{ns}})[A]$ be a query, $q$, over an attribute $A$ of the $R_s$ and $R_{\mathit{ns}}$ relations. Let $X$ be the auxiliary information about the sensitive data, and $\mathit{Pr_{Adv}}$ be the probability of the adversary knowing any information. A query execution mechanism ensures the partitioned data security if the following two properties hold:

 \begin{itemize}
 \item $\mathit{Pr}_{\mathit{Adv}}[e_i \overset{\mathrm{a}}{=} \mathit{ns}_j|X] = \mathit{Pr}_{\mathit{Adv}}[e_i \overset{\mathrm{a}}{=} \mathit{ns}_j|X, q(R_s,R_{\mathit{ns}})[A]]$, where $e_i= E(t_i)[A]$ is the encrypted representation for the attribute value $A$ for any tuple $t_i$ of the relation $R_s$ and $\mathit{ns}_j$ is a value for the attribute $A$ for any tuple of the relation $R_{\mathit{ns}}$. The notation $\overset{\mathrm{a}}{=}$ shows a sensitive value is identical to a non-sensitive value. This equation captures the fact that an initial probability of linking a sensitive tuple with a non-sensitive tuple will be identical after executing several queries on the relations.
\item $\mathit{Pr}_{\mathit{Adv}}[v_i \overset{\mathrm{r}}{\sim} v_j|X] = \mathit{Pr}_{\mathit{Adv}}[v_i \overset{\mathrm{r}}{\sim} v_j | X, q(R_s,R_{\mathit{ns}})[A]]$, for all $v_i, v_i \in \mathit{Domain}(A)$. The notation $\overset{\mathrm{r}}{\sim}$ shows a relationship between counts of the number of tuples with sensitive values. This equation states that the probability of adversary gaining information about the relative frequency of sensitive values does not increase by the query execution.
\end{itemize}

The definition above formalizes the security requirement of any partitioned computation approach. Of course, a partitioned approach, besides being secure, must also be correct in that it returns
the same answer as that returned by the original query $Q$ if it were to execute without regard to security.


\section{Query Binning: A Technique for Partitioned Computations using a Cryptographic Technique at the Public Cloud}
\label{sec:Partitioned Computations using Existing Cryptographic Mechanisms}

In this section, we will study query binning (QB) as a partitioned computing approach. QB is related to bucketization, which is studied in past~\cite{DBLP:conf/sigmod/HacigumusILM02}. While bucketization was carried over the data in~\cite{DBLP:conf/sigmod/HacigumusILM02}, QB performs  bucketization on queries. In general, one may ask more queries than original query while adding overhead but it prevents the above-mentioned inference attack. We study QB under some assumption and setting, given below.\footnote{Some of these assumptions are made primarily for ease of the exposition and will be relaxed in~\cite{TR-2018}.}.

\smallskip\noindent{\bf Problem Setup.} We assume the following two entities in our model: (\textit{i}) \emph{A database (DB) owner}: who splits each relation $R$ in the database having attributes $R_s$ and $R_{\mathit{ns}}$ containing all sensitive and non-sensitive tuples, respectively. (\textit{ii}) \emph{A public cloud}: The DB owner outsources the relation $R_{\mathit{ns}}$ to a public cloud. The tuples in $R_s$ are encrypted using any existing mechanism before outsourcing to the same public cloud. However, in the approach, we use non-deterministic encryption, \textit{i}.\textit{e}., the cipher representation of two occurrences of an identical value has different representations.

\smallskip\noindent{\bf DB Owner Assumptions.} In our setting, the DB owner has to store some (limited) metadata such as searchable values and their frequency counts, which will be used for appropriate query formulation. The DB owner is assumed to have sufficient storage for such metadata, and also computational capabilities to perform encryption and decryption. The size of metadata is exponentially smaller than the size of the original data.

\smallskip\noindent{\bf Adversarial Model.} The adversary (\textit{i}.\textit{e}., the untrusted cloud) is assumed to be honest-but-curious, which is a standard setting for security in the public cloud that is \textit{not trustworthy}. An honest-but-curious adversarial public cloud, thus, stores an outsourced dataset without tampering, correctly computes assigned tasks, and returns answers; however, it may exploit side knowledge (\textit{e}.\textit{g}., query execution, background knowledge, and the output size) to gain as much information as possible about the sensitive data. Furthermore, the adversary can eavesdrop on the communication channels between the cloud and the DB owner, and that may help in gaining knowledge about sensitive data, queries, or results. The adversary has full access to the following information: (\textit{i}) all non-sensitive data outsourced in plaintext, and (\textit{ii}) some \emph{auxiliary} information of the sensitive data. The auxiliary information may contain the metadata of the relation and the number of tuples in the relation. Furthermore, the adversary can observe frequent query types and frequent query terms on the non-sensitive data in case of selection queries. The honest-but-curious adversary, however, cannot launch any attack against the DB owner.

\smallskip\noindent\textbf{Assumptions for QB.} We develop QB initially under the assumption that queries are only on a single attribute, say $A$. The QB approach takes as inputs: (\textit{i}) the set of data values (of the attribute $A$) that are sensitive along with their counts, and (\textit{ii}) the set of data values (of the attribute $A$) that are non-sensitive, along with their counts. The QB returns a partition of attribute values that form the query bins for both the sensitive as well as for the non-sensitive parts of the query.

In this chapter, we also restrict to a case when a value has at most two tuples, where one of them must be sensitive and the other one must be non-sensitive, but both the tuples cannot be sensitive or non-sensitive. The scenario depicted in Example 1 satisfies this assumption. The \emph{EId} attribute values corresponding to sensitive tuples include $\langle$\texttt{E101}, \texttt{E259}, \texttt{E152}, \texttt{E159}$\rangle$ and from the non-sensitive relation values are $\langle$\texttt{E199}, \texttt{E259}, \texttt{E152}, \texttt{E254}$\rangle$. Note that all the values occur only one time in one set.

\smallskip\noindent\textbf{Full version.} In this chapter, we restrict the algorithm for selection query only on one attribute. The full details of the algorithm, extensions of the algorithm for values having a different number of tuples, conjunctive, range, join, insert queries, and dealing with the workload-skew attack is addressed in~\cite{TR-2018}. Further, the computing cost analysis and efficiency analysis of QB at different or identical-levels of security against a pure cryptographic technique is given in~\cite{TR-2018}.

\smallskip\noindent\textbf{The Approach.} We develop an efficient approach to execute selection queries securely (preventing the information leakage as shown in Example 1) by appropriately partitioning the query at a public cloud, where sensitive data is cryptographically secure while non-sensitive data stays in cleartext. For answering a selection query, naturally, we use any existing cryptographic technique on sensitive data and a simple search on the cleartext non-sensitive data. Naturally, we can use a secure hardware, \textit{e}.\textit{g}., Intel SGX, for all such operations; however, as mentioned in \S\ref{sec:introduction} Figure~\ref{fig:compare}, SGX-based processing takes a significant amount of time, due to limited space of the enclave.


Informally, QB distributes attribute values in a matrix, where rows are sensitive bins, and columns are non-sensitive bins. For example, suppose there are 16 values, say $0, 1,\ldots,15$, and assume all the values have sensitive and associated non-sensitive tuples. Now, the DB owner arranges 16 values in a $4\times4$ matrix, as follows:

\begin{center}
\begin{tabular}{|l|l|l|l|l|}
  \hline
             & $\mathit{NSB}_0$ & $\mathit{NSB}_1$ & $\mathit{NSB}_2$ & $\mathit{NSB}_3$  \\  \hline  \hline
    $\mathit{SB}_0$ & 11  & 2 & 5 & 14 \\  \hline
    $\mathit{SB}_1$ &  10&  3&  8&7  \\  \hline
    $\mathit{SB}_2$ &  0&  15& 6 & 4 \\  \hline
    $\mathit{SB}_3$ &  13&  1& 12 &9  \\  \hline
    \end{tabular}
\end{center}

In this example, we have four sensitive bins: $\mathit{SB}_0$ \{11,2,5,14\}, $\mathit{SB}_1$ \{10,3,8,7\}, $\mathit{SB}_2$ \{0,15,6,4\}, $\mathit{SB}_3$ \{13,1,12,9\}, and four non-sensitive bins: $\mathit{NSB}_0$ \{11,10,0,13\}, $\mathit{NSB}_1$ \{2,3,15,1\}, $\mathit{NSB}_2$ \{5,8,6,12\}, $\mathit{NSB}_3$ \{14,7,4,9\}. When a query arrives for a value, say 1, the DB owner searches for the tuples containing values 2,3,15,1 (viz. $\mathit{NSB}_1$) on the non-sensitive data and values in $\mathit{SB}_3$ (viz., 13,1,12,9) on the sensitive data using the cryptographic mechanism integrated into QB. While the adversary learns that the query corresponds to one of the four values in $\mathit{NSB}_1$, since query values in $\mathit{SB}_3$ are encrypted, the adversary does not learn any sensitive value or a non-sensitive value that is identical to a clear-text sensitive value.

\LinesNotNumbered \begin{algorithm}[!h]
\textbf{Inputs:} $|\mathit{NS}|$: the number of values in the non-sensitive data, $|S|$: the number of values in the sensitive data.

\textbf{Outputs:} $\mathit{SB}$: sensitive bins; $\mathit{NSB}$: non-sensitive bins

\nl{\bf Function $\mathit{create\_bins(S,NS)}$} \nllabel{ln:function_create_bucket}
\Begin{
\nl Permute all sensitive values \nllabel{ln:permute}

\nl $x, y \leftarrow \mathit{approx\_sq\_factors(|NS|)}$: $x \geq y$ \nllabel{ln:largest_divisors}

\nl $|\mathit{NSB}| \leftarrow x$, $\mathit{NSB} \leftarrow \lceil |\mathit{NS}|/x\rceil$, $\mathit{SB} \leftarrow x$, $|\mathit{SB}| \leftarrow y$ \nllabel{ln:number_of_buckets}

\nl \lFor{$i \in (1,|S|)$}{$\mathit{SB}[i$ modulo $x][\ast]\leftarrow S[i]$\nllabel{ln:sesnitive_allocate}}

\nl \lFor{$(i,j)\in (0,\mathit{SB}-1),(0,|\mathit{SB}|-1)$}{$\mathit{NSB}[j][i]\leftarrow \mathit{allocateNS(\mathit{SB}[i][j])}$ \nllabel{ln:assign_value_to_NS_bucket}}

\nl \lFor{$i\in (0,\mathit{NSB}-1)$}{$\mathit{NSB}[i][\ast]\leftarrow$ fill the bin if empty with the size limit to $x$ \nllabel{ln:assign_remaining_NS}}

\nl \Return $\mathit{SB}$ and $\mathit{NSB}$
}


\nl{\bf Function $\mathit{allocateNS(\mathit{SB}[i][j])}$} \nllabel{ln:function_allocateNS}
\Begin{
find a non-sensitive value associated with the $j^{\mathit{th}}$ sensitive value of the $i^{\mathit{th}}$ sensitive bin
}

\caption{Bin-creation algorithm, the base case.}
\label{alg:bin_creation}
\end{algorithm}

Formally, QB appropriately maps a selection query for a keyword $w$, say $q(w)$, to corresponding queries over the non-sensitive relation, say $q(W_{\mathit{ns}})(R_{\mathit{ns}})$, and encrypted relation, say $q(W_s)(R_s)$. The queries $q(W_{\mathit{ns}})(R_{\mathit{ns}})$ and $q(W_s)(R_s)$, each of which represents a set of query values that are executed over the relation $R_{\mathit{ns}}$ in plaintext and, respectively, over the sensitive relation $R_s$, using the underlying cryptographic method. The sets $W_{\mathit{ns}}$ from $R_{\mathit{ns}}$ and $W_s$ from $R_s$ are selected such that: (\textit{i}) $w \in q(W_{\mathit{ns}})(R_{\mathit{ns}}) \cap q(W_s)(R_s)$ to ensure that all the tuples containing $w$ are retrieved, and, (\textit{ii}) the execution of the queries $q(W_{\mathit{ns}})(R_{\mathit{ns}})$ and $q(W_s)(R_s)$ does not reveal any information (and $w$) to the adversary. The set of $q(W_{\mathit{ns}})(R_{\mathit{ns}})$ is entitled non-sensitive bins, and the set of $q(W_s)(R_s)$ is entitled sensitive bins. Algorithm~\ref{alg:bin_creation} provides pseudocode of bin-creation method.\footnote{The function $\mathit{approx\_sq\_factors}$ in Algorithm~\ref{alg:bin_creation} two factors $x$ and $y$ of a number $n$, such that either they are equal or close to each other so that the difference between $x$ and $y$ is less than the difference between any two factors of $n$ (and $x \times y =n$).} Results from the execution of the queries $q(W_{\mathit{ns}})(R_{\mathit{ns}})$ and $q(W_s)(R_s)$ are decrypted, possibly filtered, and merged to generate the final answer.

Based on QB Algorithm~\ref{alg:bin_creation}, for answering the above-mentioned three queries in Example 1, given in Section~\ref{subsec:Inference Attack and Partitioned Data Security}, Algorithm~\ref{alg:bin_creation} creates two sets or bins on sensitive parts: sensitive bin 1, denoted by $\mathit{SB}_1$, contains $\{$\texttt{E101}, \texttt{E259}$\}$, sensitive bin 2, denoted by $\mathit{SB}_2$, contains $\{$\texttt{E152}, \texttt{E159}$\}$, and two sets/bins on non-sensitive parts: non-sensitive bin 1, denoted by $\mathit{NSB}_1$, contains $\{$\texttt{E259}, \texttt{E254}$\}$, non-sensitive bin 2, denoted by $\mathit{NSB}_2$, contains $\{$\texttt{E199}, \texttt{E152}$\}$.

\begin{table}[!h]
  \centering
    \begin{tabular}{|l|l|l|}
    \hline
    Query value & \multicolumn{2}{|c|}{Returned tuples/Adversarial view}           \\ \hline
    ~                         & Employee1 & Employee2 \\ \hline
    E259                      & $\mathit{E(t_4)}$, $\mathit{E(t_1)}$       & $t_2$, $t_6$  \\ \hline
    E101                      & $\mathit{E(t_4)}$, $\mathit{E(t_1)}$       & $t_3$, $t_8$  \\ \hline
    E199                      & $\mathit{E(t_4)}$, $\mathit{E(t_1)}$       & $t_3$, $t_8$  \\ \hline
    \end{tabular}
    \caption{Queries and returned tuples/adversarial view when following QB.}
    \label{tab:answer table qb}
    \end{table}

\DontPrintSemicolon
\LinesNotNumbered \begin{algorithm}[!h]
\textbf{Inputs:} $w$: the query value.

\textbf{Outputs:} $\mathit{SB}_a$ and $\mathit{NSB}_b$: one sensitive bin and one non-sensitive bin to be retrieved for answering $w$.

\textbf{Variables:} $\mathit{found}\leftarrow$ \textbf{false}

\nl{\bf Function $\mathit{retrieve\_bins(q(w))}$} \nllabel{ln:function_retrieve_bucket}
\Begin{

\nl \For{$(i,j) \in (0,\mathit{SB}-1),(0,|\mathit{SB}|-1)$\nllabel{ln:check_each_sbucket}}{\If{$w=\mathit{SB}_i[j]$}{

\nl \Return $\mathit{SB}_i$ and $\mathit{NSB}_j$; $\mathit{found} \leftarrow$ \textbf{true}; \textbf{break} \nllabel{ln:retrieve_rule_svalue}
}}

\nl \If{$\mathit{found} \neq$ \textbf{\textnormal{\textbf{true}}}\nllabel{ln:check_nsbuckets}}{
\nl \For{$(i,j) \in (0,\mathit{NSB}-1),(0,|\mathit{NSB}|-1)$}{

\nl \If{$w=\mathit{NSB}_i[j]$}{\Return $\mathit{NSB}_i$ and $\mathit{SB}_j$; \textbf{break}\nllabel{ln:retrieve_rule_nsvalue}}}}

\nl Retrieve the desired tuples from the cloud by sending encrypted values of the bin $\mathit{SB}_i$ (or $\mathit{SB}_j$) and clear-text values of the bin $\mathit{NSB}_j$ (or $\mathit{NSB}_i$) to the cloud

\nllabel{ln:retrieve_bin}

}
\caption{Bin-retrieval algorithm.}
\label{alg:bin_retrieval}
\end{algorithm}

Algorithm~\ref{alg:bin_retrieval} provides a way to retrieve the bins. Thus, by following Algorithm~\ref{alg:bin_retrieval}, Table~\ref{tab:answer table qb} shows that the adversary cannot know the query value $w$ or find a value that is shared between the two sets, when answering to the above-mentioned three queries. The reason is that the desired query value, $w$, is encrypted with other encrypted values of the set $W_s$, and, furthermore, the query value, $w$, is obscured in many requested non-sensitive values of the set $W_{\mathit{ns}}$, which are in cleartext. Consequently, the adversary is unable to find an intersection of the two sets, which is the exact value. Thus,
while answering a query, the adversary cannot learn which employee works only in defense, design, or in both.

\smallskip\noindent\textbf{Correctness.} The correctness of QB indicates that the approach maintains
an initial probability of \emph{associating} a sensitive tuple with a non-sensitive tuple will be identical after executing several queries on the relations. 

We can illustrate the correctness of QB with the help of an example. The objective of the adversary is to deduce a clear-text value corresponding to an encrypted value of either $\{$\texttt{E101, E259}$\}$ or $\{$\texttt{E152, E159}$\}$, since we retrieve the set of these two values. Note that before executing a query, the probability of an encrypted value, say $E_i$, (where $E_i$ may be \texttt{E101}, \texttt{E259}, \texttt{E152}, or \texttt{E159}) to have the clear-text value is 1/4, which QB maintains at the end of a query. Assume that $E_1$ and $E_2$ are encrypted representations of \texttt{E101} and \texttt{E259}, respectively. Also, assume that $v_1$, $v_2$, $v_3$, $v_4$ are showing the cleartext value of \texttt{E259}, \texttt{E254}, \texttt{E199}, and \texttt{E152}, respectively.

When the query arrives for $\langle E_1,E_2,v_1,v_2\rangle$, the adversary gets the fact that the clear-text representation of $E_1$ and $E_2$ cannot be $v_1$ and $v_2$ or $v_3$ and $v_4$. If this will happen, then there is no way to associate each sensitive bin of the new bipartite graph with each non-sensitive bin. Now, if the adversary considers the clear-text representation of $E_1$ is $v_1$, then the adversary have four possible allocations of the values $v_1$, $v_2$, $v_3$, $v_4$ to $E_1$, $E_2$, $E_3$, $E_4$, such as:
$\langle v_1,v_2,v_3,v_4\rangle$,
$\langle v_1,v_2,v_4,v_3\rangle$,
$\langle v_1,v_3,v_4,v_2\rangle$,
$\langle v_1,v_4,v_3,v_2\rangle$.

Since the adversary is not aware of the exact clear-text value of $E_1$, the adversary also considers the clear-text representation of $E_1$ is $v_2$, $v_3$, or $v_4$. This results in 12 more possible allocations of the values $v_1$, $v_2$, $v_3$, $v_4$ to $E_1$, $E_2$, $E_3$, $E_4$. Thus, the retrieval of the four tuples containing one of the following: $\langle E_1,E_2,v_1,v_2\rangle$, results in 16 possible allocations of the values $v_1$, $v_2$, $v_3$, and $v_4$ to $E_1$, $E_2$, $E_3$, and $E_4$, of which only four possible allocations have $v_1$ as the clear-text representation of $E_1$. This results in the probability of finding $E_1=v_1$ is 1/4.

\begin{figure}[h!]
\centering
\includegraphics[scale=0.7]{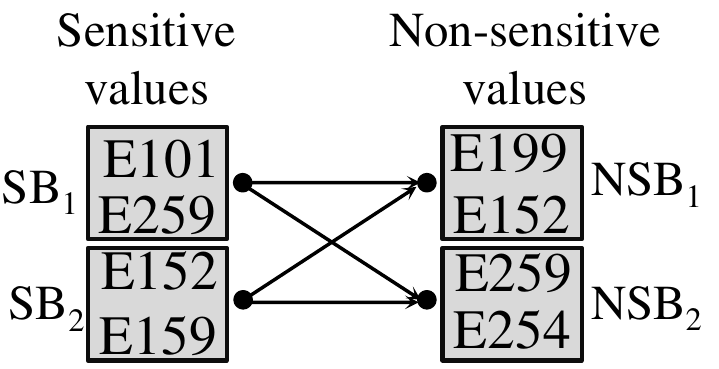}
  \caption{A bipartite graph showing sensitive and non-sensitive bins after query execution, where each sensitive value gets associated with each non-sensitive value.}
  \label{fig:survival_matching2}
\end{figure}

Note that following this technique, executing queries under for each keyword will not eliminate any surviving matches of the bipartite graph, and hence, the adversary can find the new bipartite graph identical to a bipartite graph before the query execution. Figure~\ref{fig:survival_matching1} shows an initial bipartite graph before the query execution and Figure~\ref{fig:survival_matching2} shows a bipartite graph after the query execution when creating bins on the values. Note that in Figure~\ref{fig:survival_matching2} each sensitive bin is linked to each non-sensitive bin, that in turns, shows that each sensitive value is linked to each non-sensitive value.

\section{Effectiveness of QB}
\label{sec:Comparing QB with a Cryptographic Approach}

From the performance perspective, QB results in saving of encrypted data processing over non-sensitive data -- the more the non-sensitive data, the more potential savings. Nonetheless, QB incurs overhead -- it converts a single predicate selection query into a set of predicates selection queries over cleartext non-sensitive data, and, a set of encrypted predicates selection queries albeit over a smaller database consisting only of sensitive data. In this section, we compare QB against a pure cryptographic technique and show when using QB is beneficial.

For our model, we will need the following notations: (\textit{i}) $C_{\mathit{com}}$: Communication cost of moving one tuple over the network. (\textit{ii}) $C_p$ (or $C_e$): Processing cost of a single selection query on plaintext (or encrypted data). In addition, we define three parameters:
\begin{description}
  \item [$\alpha$]: is the ratio between the sizes of the sensitive data (denoted by $S$) and the entire dataset (denoted by $S+\mathit{NS}$, where $\mathit{NS}$ is non-sensitive data).

  \item [$\beta$]: is the ratio between the predicate search time on encrypted data using a cryptographic technique and on clear-text data. The parameter $\beta$ captures the overhead of a cryptographic technique. Note that $\beta = C_e/C_p$. 

  \item [$\gamma$]: is the ratio between the processing time of a single selection query on encrypted data and the time to transmit the single tuple over the network from the cloud to the DB owner. Note that $\gamma = C_e/C_{\mathit{com}}$. 
\end{description}

Based on the above parameters, we can compute the cost of cryptographic and non-cryptographic selection operations as follows:
\begin{description}
  \item[$\mathit{Cost}_{\mathit{plain}}(x, D)$] is the sum the processing cost of $x$ selection queries on plaintext data and the communication cost of moving all the tuples having $x$ predicates from the cloud to the DB owner, \textit{i}.\textit{e}., $x(\log(D)P_p+ \rho D C_{\mathit{com}})$.

  \item[$\mathit{Cost}_{\mathit{crypt}}(x, D)$] is the sum the processing cost of $x$ selection queries on encrypted data and the communication cost of moving all the tuples having $x$ predicates from the cloud to the DB owner, i.e., $P_e D+ \rho x D C_{\mathit{com}}$, where $\rho$ is the selectivity of the query. Note that cost of evaluating $x$ queries over encrypted data using techniques such as~\cite{DBLP:conf/sp/SongWP00,DBLP:conf/eurocrypt/GilboaI14,DBLP:journals/isci/EmekciMAA14}, is amortized and can be performed using a single scan of data. Hence, $x$ is not the factor in the cost corresponding to encrypted data processing.
\end{description}

Given the above, we define a parameter $\eta$ that is the ratio between the computation and communication cost of searching using QB and the computation and communication cost of searching when the entire data (viz. sensitive and non-sensitive data) is fully encrypted using the cryptographic mechanism.
$$\eta = \frac{\mathit{Cost}_{\mathit{crypt}}(|\mathit{SB}|, S)}{\mathit{Cost}_{\mathit{crypt}}(1,D)} + \frac{\mathit{Cost}_{\mathit{plain}}(|\mathit{NSB}|, \mathit{NS})}{\mathit{Cost}_{\mathit{crypt}}(1,D)}$$
\noindent Filling out the values from above, the ratio is:
$$\eta=\frac{C_e S + |\mathit{SB}|\rho D C_{\mathit{com}}}{C_e D + \rho D C_{\mathit{com}}} + \frac{|\mathit{NSB}|\log(D)C_p  + |\mathit{NSB}| \rho D C_{\mathit{com}}}{C_e D + \rho D C_{\mathit{com}}}$$
\noindent Separating out the communication and processing costs, $\eta$ becomes:
$$\eta=\frac{S}{D}\frac{C_e}{C_e + \rho C_{\mathit{com}}} + \frac{|\mathit{NSB}|\log(D)C_p}{C_e D + \rho D C_{\mathit{com}}}
+\frac{\rho D C_{\mathit{com}}(|\mathit{NSB}|+|\mathit{SB}|)}{C_e D + \rho D C_{\mathit{com}}}$$
\noindent Substituting for various terms and cancelling common terms provides:
$$\eta = \alpha \frac{1}{(1 + \frac{\rho}{\gamma})} + \frac{\log(D)}{D} \frac{|NSB|}{\beta (1 + \frac{\rho}{\gamma})} + \frac{\rho}{\gamma} \frac{|\mathit{NSB}| + |\mathit{SB}|}{(1 + \frac{\rho}{\gamma})}$$
Note that $\rho/\gamma$ is very small, thus the term $(1 + \rho/\gamma)$ can be substituted by $1$. Given the above, the equation becomes:
$$\eta = \alpha  + \log(D)|\mathit{NSB}/D\beta + \rho(|\mathit{NSB}| + |\mathit{SB}|)/\gamma$$
Note that the term $\log(D)|\mathit{NSB}|/D\beta$ is very small since $|\mathit{NSB}|$ is the number of distinct values (approx. equal to $\sqrt{|\mathit{NS}|}$) in a non-sensitive bin, while
$D$, which is the size of a database, is a large number, and $\beta$ value is also very large. Thus, the equation becomes: $$\eta = \alpha  + \rho(|\mathit{SB}|+ |\mathit{NSB}|)/\gamma$$

QB is better than a cryptographic approach when $\eta < 1$, \textit{i}.\textit{e}., $\alpha  + \rho(|\mathit{SB}|+ |\mathit{NSB}|)/\gamma < 1$. Thus, $$\alpha   < 1 -  \frac{\rho(|\mathit{SB}|+ |\mathit{NSB}|)}{\gamma}$$ Note that the values of $|\mathit{SB}| $ and $|\mathit{NSB}|$ are approximately $\sqrt{|\mathit{NS}|}$, we can simplify the above equation to: $\alpha   < 1 -  2 \rho \sqrt{|\mathit{NS}|}/\gamma$. If we estimate $\rho$ to be roughly $1/|\mathit{NS}|$ (i.e., we assume uniform distribution), the above equation becomes: $\alpha  < 1 -  2/\gamma \sqrt{|\mathit{NS}|}$.

\begin{figure}[h]
							\centering
				\includegraphics[scale=0.55]{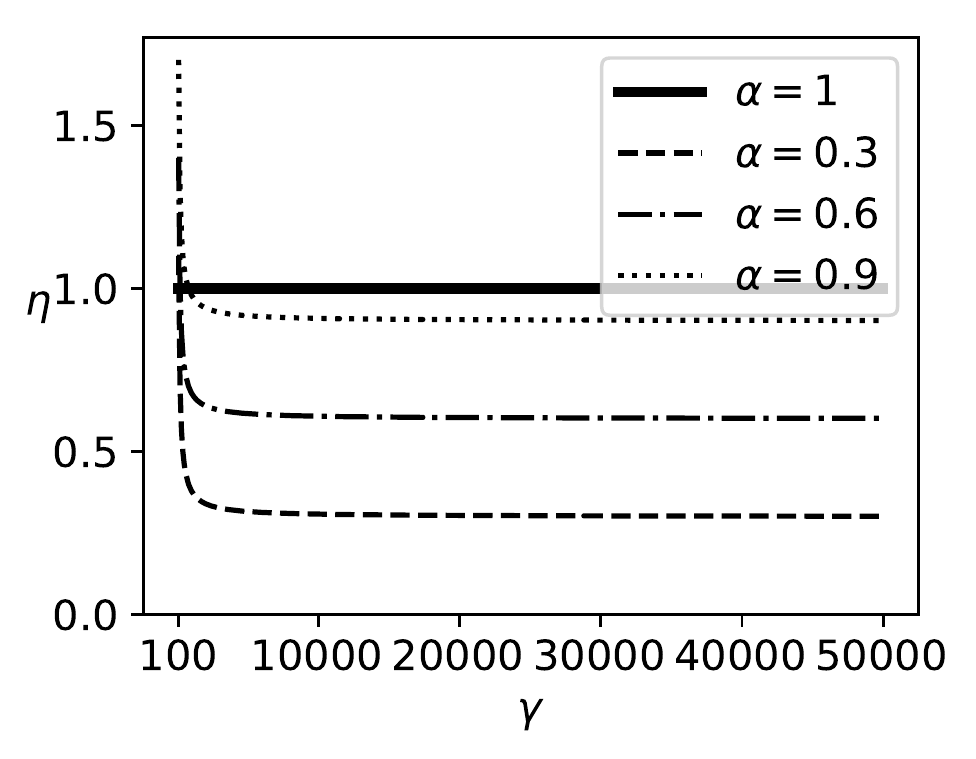}

\caption{Efficiency graph using equation $\eta = \alpha  + \rho(|\mathit{SB}|+ |\mathit{NSB}|)/\gamma$.}
				\label{fig:non-index-eta-graph-using-equation}
\end{figure}

The equation above demonstrates that QB trades increased communication costs to reduce the amount of data that needs to be searched in encrypted form. Note that the reduction in encryption cost is proportional to $\alpha$ times the size of the database, while the increase in communication costs is proportional to $\sqrt{|D|}$, where $|D|$ is the number of distinct attribute values. This, coupled with the fact that $\gamma$ is much higher than 1 for encryption mechanisms that offer strong security guarantees, ensures that QB almost always outperforms the full encryption approaches. For instance, the cryptographic cost for search using secret-sharing is $\approx 10ms$ \cite{DBLP:journals/isci/EmekciMAA14}, while the cost of transmitting a single row ($\approx $ 200 bytes for TPCH Customer table) is $\approx 4$ $\mu$$s$ making the value of $\gamma \approx 25000$. Thus, QB, based on the model, should outperform the fully encrypted solution for almost any value of $\alpha$, under ideal situations where our assumption of uniformity holds. Figure~\ref{fig:non-index-eta-graph-using-equation} plots a graph of $\eta$ as a function of $\gamma$, for varying sensitivity and $\rho = 10\%$.

\begin{figure}			
	\centering
				\includegraphics[scale=0.55]{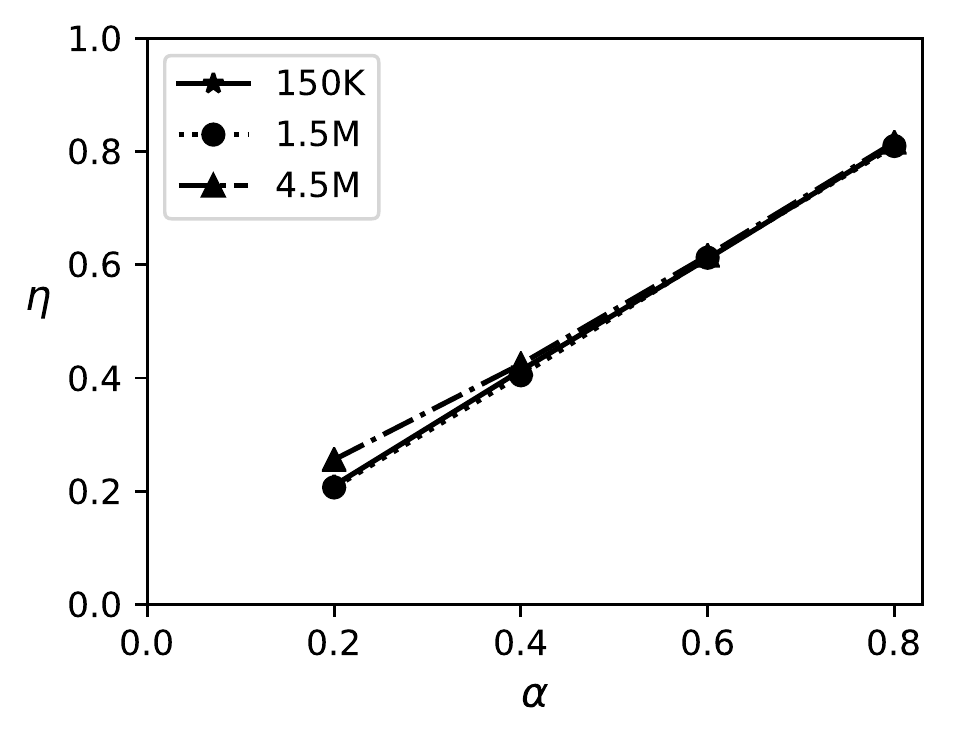}
				\caption{Dataset size.}
				\label{fig:data_set_size_increase}
	\end{figure}

Figure~\ref{fig:data_set_size_increase} plots $\eta$ values for the three DB sizes: 150K, 1.5M, and 4.5M for a linear scan cryptographic technique while varying sensitivity, denoted by $\alpha$. The figure shows that $\eta < 1$, irrespective of the DB sizes, confirming that the approach scales to larger DB sizes, and we do not pay any overhead while using linear-scan techniques. Recall that the reason is that the cost of searching a single value over encrypted data using a linear scan absorbs the cost of searching multiple values using a linear scan on the same data.

\section{Conclusion}
This chapter focuses on partitioned computing as a mechanism to scale cryptographic techniques  while still ensuring security. Inspired  by current security practice wherein organizations classify data as sensitive/non-sensitive prior to outsourcing, partitioned computing exploits the fact that non-sensitive data can be outsourced to and processed  at the public cloud  in plaintext. Thus, computation is partitioned into two parts --- the one that executes non-sensitive data in plaintext and the part that executes on sensitive data. We study partitioned computing in two different settings --- in the context of the hybrid cloud wherein sensitive data is kept on a private cloud and in the context of a public cloud where sensitive data is encrypted using existing cryptographic mechanisms prior to outsourcing. In both cases, partitioned computing if not done carefully could leak sensitive data. We define a notion of partitioned security (primarily for the context of partitioned computing in the public cloud)  and develop mechanisms to ensure that partitioned computing does not lead to any new vulnerabilities.

\end{document}